\documentclass[12pt]{article}
\usepackage{amssymb,amsmath,epsfig}
\begin{document}
\title{\bf Energy Distribution in $f(R)$ Gravity}

\author{M. Sharif \thanks{msharif@math.pu.edu.pk} and M. Farasat
Shamir \thanks{frasat@hotmail.com}\\\\
Department of Mathematics, University of the Punjab,\\
Quaid-e-Azam Campus, Lahore-54590, Pakistan.}

\date{}

\maketitle
\begin{abstract}
The well-known energy problem is discussed in $f(R)$ theory of
gravity. We use the generalized Landau-Lifshitz energy-momentum
complex in the framework of metric $f(R)$ gravity to evaluate the
energy density of plane symmetric solutions for some general
$f(R)$ models. In particular, this quantity is found for some
popular choices of $f(R)$ models. The constant scalar curvature
condition and the stability condition for these models are also
discussed. Further, we investigate the energy distribution of
cosmic string spacetime.
\end{abstract}

{\bf Keywords:}  $f(R)$ gravity, Generalized Landau-Lifshitz EMC,
Energy Density.

\section{Introduction}

The energy localization has been a thorny problem since the Einstein
era. Several attempts have been made to find a general and unique
tensor representation for the energy-momentum. Einstein was the
first who tried to solve this problem by introducing energy-momentum
pseudo tensors. He established the energy-momentum conservation laws
given by \cite{1}
\begin{equation*}
\frac{\partial}{\partial
x^\nu}\{\sqrt{-g}(T_\mu^\nu+t_\mu^\nu)\}=0,\quad
(\mu,\nu=0,1,2,3),
\end{equation*}
where $T_\mu^\nu$ is the energy-momentum density of matter and
$t_\mu^\nu$ represents the energy-momentum density of gravitation.
It is mentioned here that $t_\mu^\nu$ is not a tensor quantity
rather it is the gravitational field pseudo-tensor. Komar \cite{01}
gave a set of energy-momentum covariant conservation laws and
developed their relationship to the generators of infinitesimal
coordinate transformations. Bergmann \cite{02}-\cite{04} contributed
greatly to the fundamental nature of conservation laws.

Landau-Lifshitz introduced \cite{2} the energy-momentum complex by
using the geodesic coordinate system at some particular point of
space. Many other people like Tolman \cite{3}, Papapetrou \cite{4},
Bergmann \cite{5}, Goldberg \cite{6}, M$\varnothing$ller \cite{7}
and Weinberg \cite{8} developed their own energy-momentum complexes.
All these prescriptions, except M$\varnothing$ller, are restricted
to perform calculations in Cartesian coordinates only. Also, we
cannot define angular momentum with the help of these prescriptions.
This idea of energy-momentum pseudo-tensors was severely criticized
by some people. Even it was quoted in a famous book \cite{9} that:
\textit{Anyone who seeks for a general formula for "local
gravitational energy-momentum" is asking for the right answer to the
wrong question}. Misner et al. \cite{9} showed that energy can only
be localized in spherical systems. But later on, Cooperstock and
Sarracino \cite{10} proved that if energy is localizable for
spherical systems, then it can be localized in any system. Bondi
\cite{11} argued that a non-localizable form of energy is not
allowed in General Relativity.

The idea of quasi-local energy was proposed by some authors
\cite{12}-\cite{15}. In this method , we can use any coordinate
system while finding the quasi-local masses to obtain the
energy-momentum of a curved spacetime. Chang et al. \cite{16} proved
that every energy-momentum complex can be associated with a
particular Hamiltonian boundary term. Thus the energy-momentum
complexes may also be considered as quasi-local. Virbhadra and his
collaborators \cite{17} verified for asymptotically flat spacetimes
that different energy-momentum complexes could give the same result
for a given spacetime. They also found encouraging results for the
case of asymptotically non-flat spacetimes by using different
energy-momentum complexes. Senovilla \cite{18} constructed
super-energy tensors for arbitrary fields in any dimension. These
tensors had good mathematical and physical properties, and in
general, the completely timelike component of these super-energy
tensors had the mathematical features of an energy density.

It might be interesting if this problem could be explored in the
alternative theories of gravity. Recently, some work about
energy-momentum has been investigated in teleparallel theory of
gravity \cite{21}-\cite{23} with the hope that this problem may be
settled down in this theory. For this purpose, the teleparallel
versions of M$\varnothing$ller, Bergmann, Einstein and
Landau-Lifshitz prescriptions are derived. Sharif and Jamil
\cite{24} used these prescriptions to explore the energy-momentum
distribution for particular spacetimes. It is concluded that results
are consistent in some cases but no general conclusion can be
deduced.

The $f(R)$ theory of gravity is another alternative theory of
gravity which has received much attention in recent years due to its
cosmologically important $f(R)$ models. These models include higher
order curvature invariants as function of Ricci scalar. It has been
shown \cite{024}-\cite{026} that some $f(R)$ models pass solar
system test. In particular, Nojiri and Odintsov \cite{024} proposed
$f(R)$ models with negative and positive powers of the curvature. It
is shown that the terms with positive powers of the curvature
provide the inflationary epoch while the terms with negative powers
of the curvature serve as effective dark energy, supporting current
cosmic acceleration. They also discussed the consistency of some
$f(R)$ models which include the terms involving logarithm of scalar
curvature. Cognola et al. \cite{027} introduced a class of
exponential $f(R)$ models. They proved that these models passed all
local tests, including stability of spherical body solution,
non-violation of Newton's law, and generation of a very heavy
positive mass for the additional scalar degree of freedom. Amendola
et al. \cite{028} derived the conditions under which dark energy
$f(R)$ models are cosmologically viable. Thus $f(R)$ theory of
gravity seems attractive due to cosmologically important $f(R)$
models. It is hoped that the issue of energy-momentum localization
can be addressed in this theory.

In a recent paper, Multam$\ddot{a}$ki et al. \cite{25} studied
energy-momentum complexes in this theory. They generalized the
Landau-Lifshitz prescription to calculate energy-momentum in the
framework of metric $f(R)$ gravity. As an important special case,
they evaluated the energy density for the Schwarzschild de Sitter
spacetime. Bertolami and Sequeira \cite{26} discussed some $f(R)$
models and studied them from the point of view of the energy
conditions and of their stability under the Dolgov-Kawasaki
criterion.

In this paper, we investigate energy distribution of some static
plane symmetric solutions \cite{27} using the generalized
Landau-Lifshitz energy momentum complex. We also explore energy
density of cosmic string spacetime. These results are also found
for some important $f(R)$ models. The stability and constant
scalar curvature conditions of these models are also discussed.
The paper is organized as follows: In section \textbf{2}, we give
a brief introduction about the field equations and the generalized
Landau-Lifshitz energy-momentum complex in the context of metric
$f(R)$ gravity.  In sections \textbf{3} and \textbf{4}, the energy
distribution of plane symmetric solutions and cosmic string
spacetime are found respectively using the generalized
Landau-Lifshitz energy-momentum complex. In the last section, we
summarize and conclude the results.

\section{Generalized Landau-Lifshitz Energy-Momentum Complex}

The $f(R)$ theory of gravity modifies or generalizes the general
theory of relativity. The action for $f(R)$ gravity is
\begin{equation}\label{1}
S=\int\sqrt{-g}(\frac{1}{16\pi{G}}f(R)+L_{m}).
\end{equation}
Here $f(R)$ is a general function of the Ricci scalar. We note that
this action is obtained by replacing $R$ with $f(R)$ in the standard
Einstein-Hilbert action. The corresponding field equations are found
by varying this action with respect to the metric tensor
$g_{\mu\nu}$
\begin{equation}\label{2}
F(R)R_{\mu\nu}-\frac{1}{2}f(R)g_{\mu\nu}-\nabla_{\mu}
\nabla_{\nu}F(R)+g_{\mu\nu} \Box F(R)=\kappa T_{\mu\nu},
\end{equation}
where
\begin{equation}\label{3}
F(R)\equiv df(R)/dR,\quad\Box\equiv\nabla^{\mu}\nabla_{\mu}
\end{equation}
and $\nabla_{\mu}$ represent the covariant derivative. Contracting
the field equations, we get
\begin{equation}\label{4}
F(R)R-2f(R)+3\Box F(R)=\kappa T
\end{equation}
and in vacuum, this reduces to
\begin{equation}\label{5}
F(R)R-2f(R)+3\Box F(R)= 0.
\end{equation}
Equation (\ref{5}) gives a relationship between $f(R)$ and $F(R)$.
This equation shows that any metric with constant scalar curvature,
say $R=R_{0}$, is a solution of the contracted equation (\ref{5}) as
long as the following equation holds
\begin{equation}\label{6}
F(R_{0})R_{0}-2f(R_{0})=0.
\end{equation}
This gives the condition of constant scalar curvature. For
non-vacuum case, this is given by
\begin{equation}\label{06}
F(R_{0})R_{0}-2f(R_{0})=\kappa T.
\end{equation}
These conditions are very important because these are used to check
the acceptability of $(R)$ models. This assumption of constant
scalar curvature was firstly used by Cognola et al. \cite{029} to
investigate the solutions in $f(R)$ gravity.

The generalized Landau-Lifshitz energy-momentum complex (EMC) is
give by \cite{25}
\begin{equation}\label{15}
\tau^{\mu\nu}=f'(R_{0})\tau^{\mu\nu}_{LL}+\frac{1}{6\kappa}
(f'(R_{0})R_{0}-f(R_{0}))\frac{\partial}{\partial
x^{\lambda}}(g^{\mu\nu}x^{\lambda}-g^{\mu\lambda}x^{\nu}),
\end{equation}
where $\tau^{\mu\nu}_{LL}$ is the Landau-Lifshitz EMC evaluated in
the framework of General Relativity and $\kappa=8\pi G$. Its
$00$-component turns out to be
\begin{equation}\label{16}
\tau^{00}=f'(R_{0})\tau^{00}_{LL}+\frac{1}{6\kappa}(f'(R_{0})
R_{0}-f(R_{0}))(\frac{\partial}{\partial x^{i}}g^{00}
x^{i}+3g^{00}),
\end{equation}
where $\tau^{00}_{LL}$ is
\begin{equation}\label{17}
\tau^{00}_{LL}=(-g)(T^{00}+t^{00}_{LL})
\end{equation}
and $t^{00}_{LL}$ can be obtained from the following expression
\begin{eqnarray}\label{18}
t^{\mu\nu}_{LL}&=&\frac{1}{2\kappa}[(2\Gamma^{\gamma}_{\alpha\beta}
\Gamma^{\delta}_{\gamma\delta}-\Gamma^{\gamma}_{\alpha\delta}
\Gamma^{\delta}_{\beta\gamma}-\Gamma^{\gamma}_{\alpha\gamma}
\Gamma^{\delta}_{\beta\delta})(g^{\mu\alpha}g^{\nu\beta}
-g^{\mu\nu}g^{\alpha\beta})\nonumber\\
&+&g^{\mu\alpha}g^{\beta\gamma}(\Gamma^{\nu}_{\alpha\delta}
\Gamma^{\delta}_{\beta\gamma}+\Gamma^{\nu}_{\beta
\gamma}\Gamma^{\delta}_{\alpha\delta}
-\Gamma^{\nu}_{\gamma\delta}\Gamma^{\delta}_{\alpha\beta}
-\Gamma^{\nu}_{\alpha\beta}\Gamma^{\delta}_{\gamma \delta})\nonumber\\
&+&g^{\nu\alpha}g^{\beta\gamma}(\Gamma^{\mu}_{\alpha\delta}
\Gamma^{\delta}_{\beta\gamma}+\Gamma^{\mu}_{\beta\gamma}\Gamma^
{\delta}_{\alpha\delta}-\Gamma^{\mu}_{\gamma\delta}\Gamma^{\delta}_{\alpha\beta}
-\Gamma^{\mu}_{\alpha\beta}\Gamma^{\delta}_{\gamma\delta})\nonumber\\
&+&g^{\alpha\beta}g^{\gamma\delta}(\Gamma^{\mu}_{\alpha
\gamma}\Gamma^{\nu}_{\beta\delta}-\Gamma
^{\mu}_{\alpha\beta}\Gamma^{\nu}_{\gamma\delta})].
\end{eqnarray}
It is mentioned here that Eq.(\ref{15}) is the generalized formula
of Landau-Lifshitz energy-momentum complex valid for constant
scalar curvature. It would be worthwhile to mention here that we
need cartesian coordinates to use this formula as some energy
momentum pseudo-tensors are calculated in cartesian coordinates
only.

\section{Energy Distribution of Plane Symmetric Solutions}

This section is used to evaluate energy density of some plane
symmetric solutions found in $f(R)$ gravity \cite{27}. For this
purpose, we use the generalized Landau-Lifshitz energy-momentum
complex valid for spacetimes that have constant scalar curvature.

\subsection{Energy Density of the 1st Solution}

The first vacuum solution (Taub's metric) is given by
\begin{equation}\label{19}
ds^{2}=k_1x^{-\frac{2}{3}}dt^{2}-dx^{2}-k_2x^{\frac{4}{3}}(dy^{2}+dz^{2}),
\end{equation}
where $k_1$ and $k_2$ are arbitrary constants. The corresponding
00-component takes the form
\begin{equation}\label{20}
\tau^{00}=f'(R_{0})\tau^{00}_{LL}+\frac{11}{18\kappa k_1}
(f'(R_{0})R_{0}-f(R_{0}))(x)^{\frac{2}{3}}
\end{equation}
whereas $\tau^{00}_{LL}$ becomes
\begin{equation}\label{21}
\tau^{00}_{LL}=-\frac{1}{\kappa}(\frac{5k_2x^{\frac{1}{3}}}{3}).
\end{equation}
Thus the $00$-component of the generalized Landau-Lifshitz EMC
becomes
\begin{equation}\label{22}
\tau^{00}_{LL}=\frac{-5k_2x^{\frac{1}{3}}}{3\kappa}f'(R_{0})
+\frac{11}{18\kappa k_1} (f'(R_{0})R_{0}-f(R_{0}))(x)^{\frac{2}{3}}.
\end{equation}
Now we use $f(R)$ model to evaluate this component. It is
mentioned here that we have some restrictions for the choice of
$f(R)$ model when $R=0$. For example, we cannot use a model
including a logarithmic function of the Ricci scalar and also a
model which is a linear superposition of $R^{-m}$, where $m$ is
any positive integer. Thus we take the $f(R)$ model \cite{031} as
follows
\begin{equation}\label{23}
f(R)=R+\epsilon R^2,
\end{equation}
where $\epsilon$ is any positive real number. Consequently, the
$00$-component of the generalized Landau-Lifshitz EMC reduces to
\begin{equation}\label{24}
\tau^{00}=\frac{-5k_2x^{\frac{1}{3}}}{3\kappa}.
\end{equation}
Further, the stability condition \cite{32},
$\frac{1}{\epsilon(1+2\epsilon R_0)}=\frac{1}{\epsilon}>0$, for
the solution is satisfied.

\subsection{Energy Density of the 2nd Solution}

The second vacuum solution is
\begin{equation*}
ds^{2}=(bx+bc)^2dt^{2}-dx^{2}-e^{2a}(dy^{2}+dz^{2}).
\end{equation*}
The corresponding 00-component is
\begin{equation}\label{25}
\tau^{00}=f'(R_{0})\tau^{00}_{LL}+\frac{1}{6\kappa}
(f'(R_{0})R_{0}-f(R_{0}))(\frac{2x}{(x+c)^3}+\frac{3}{b^2(x+c)^2})
\end{equation}
while $\tau^{00}_{LL}$ becomes
\begin{equation}\label{26}
\tau^{00}_{LL}=\frac{1}{\kappa}(\frac{-e^{4a}}{(x+c)^2}).
\end{equation}
Thus the $00$-component of the generalized Landau-Lifshitz EMC turns
out to be
\begin{equation}\label{27}
\tau^{00}=\frac{1}{\kappa}(\frac{-e^{4a}}{(x+c)^2})f'(R_{0})+\frac{1}{6\kappa}
(f'(R_{0})R_{0}-f(R_{0}))(\frac{2x}{(x+c)^3}+\frac{3}{b^2(x+c)^2}).
\end{equation}
For a particulat $f(R)$ model,
\begin{equation}\label{28}
f(R)=R+\epsilon R^2,
\end{equation}
this reduces to
\begin{equation}\label{29}
\tau^{00}=\frac{1}{\kappa}(\frac{-e^{4a}}{(x+c)^2}).
\end{equation}

\subsection{Energy Density of the 3rd Solution}

The third solution ($R\neq0$) corresponds to anti deSitter metric
and is given by
\begin{equation*}
ds^{2}=e^{2(c_1x+c_2)}(dt^{2}-dy^{2}-dz^{2})-dx^{2}.
\end{equation*}
Here the 00-component becomes
\begin{equation}\label{30}
\tau^{00}=f'(R_{0})\tau^{00}_{LL}+\frac{1}{6\kappa}
(f'(R_{0})R_{0}-f(R_{0}))(\frac{3-2c_1x}{e^{2(c_1x+c_2)}})
\end{equation}
and also it follows that
\begin{equation}\label{31}
\tau^{00}_{LL}=\frac{1}{\kappa}(-5{c_1}^2e^{4(c_1x+c_2)}).
\end{equation}
Thus we obtain
\begin{equation}\label{32}
\tau^{00}=\frac{1}{\kappa}(-5{c_1}^2e^{4(c_1x+c_2)})f'(R_{0})+\frac{1}{6\kappa}
(f'(R_{0})R_{0}-f(R_{0}))(\frac{3-2c_1x}{e^{2(c_1x+c_2)}}).
\end{equation}
Now we discuss an important $f(R)$ model given by \cite{024}
\begin{equation}\label{33}
f(R)=R-\frac{a}{R}-bR^2.
\end{equation}
For $R\equiv R_0=12{c_1}^2$, we have
\begin{equation}\label{34}
f(R_0)=12{c_1}^2-\frac{a}{12{c_1}^2}-144b{c_1}^4.
\end{equation}
Inserting this value and its derivative in Eq.(\ref{32}), we get
\begin{equation}\label{35}
\tau^{00}=\frac{1}{\kappa}(-5{c_1}^2e^{4(c_1x+c_2)})(1+\frac{a}{144{c_1}^2}-24b{c_1}^2)
+\frac{1}{6\kappa}(\frac{a}{6{c_1}^2}-144b{c_1}^4)(\frac{3-2c_1x}{e^{2(c_1x+c_2)}}).
\end{equation}

It is mentioned here that this $f(R)$ model satisfy the constant
scalar condition, i.e. $F(R_0)R_0-2f(R_0)=0$ which implies that
$a=48{c_1}^4$. The stability condition \cite{25}, i.e.
$f''(R_0)\leq0$ yields
\begin{equation*}
a+b(R_0)^3\geq 0.
\end{equation*}
Since $R_0=12{c_1}^2$ and $a=48{c_1}^4$, it follows that
\begin{equation}\label{37}
1+36b{c_1}^2\geq 0.
\end{equation}
Thus the model is acceptable.

\section{Energy Distribution of Cosmic String Spacetime}

The idea of big bang suggests that universe has expanded from a hot
and dense initial condition at some finite time in the past. It is a
general cosmological assumption that the universe has gone through a
number of phase transitions at early stages of its evolution. The
cosmic string spacetime has received serious attention in recent
years due to their cosmological implications. Here we discuss energy
distribution for this cosmological model \cite{28} in $f(R)$
gravity.

Consider the non-static cosmic string spacetime \cite{28}
\begin{equation}\label{038}
ds^2=dt^2-e^{2\sqrt{\frac{\Lambda}{3}}t}[d\rho^2+(1-4GM)^2\rho^2d\phi^2+dz^2].
\end{equation}
We write this metric in Cartesian coordinates as it is required for
the generalized Landau-Lifshitz EMC. In Cartesian coordinates, this
becomes
\begin{eqnarray}\label{38}
ds^{2}&=&dt^2-e^{\alpha t}\frac{x^2+a^2y^2}{x^2+y^2}dx^2-e^{\alpha
t}\frac{y^2+a^2x^2}{x^2+y^2}dy^2\nonumber\\
&-&e^{\alpha t}dz^2+2e^{\alpha t}\frac{a^2-1}{x^2+y^2}xydxdy,
\end{eqnarray}
where $a=1-4GM$ with $G$ as the gravitational constant and $M$ as
mass per unit length of the string in the $z$ direction and
$\alpha=2\sqrt{\frac{\Lambda}{3}}$ with $\Lambda$ as the
cosmological constant. Also, the energy-momentum tensor is defined
as
\begin{equation}\label{39}
T^\nu_\mu=M\delta(x)\delta(y)diag(1,0,0,1).
\end{equation}
The Ricci scalar for this spacetime becomes
\begin{equation}\label{40}
R=-3\alpha^2=-4\Lambda.
\end{equation}
Since the Ricci scalar is constant, we can find energy density of
this model by using the generalized Landau-Lifshitz EMC. Its
$00$-component becomes
\begin{equation}\label{41}
\tau^{00}=F(R_{0})\tau^{00}_{LL}+\frac{1}{2\kappa}(F(R_{0})R_{0}-f(R_{0}))
(g^{00}).
\end{equation}

We can evaluate $\tau^{00}_{LL}$ by using
\begin{equation}\label{42}
\tau^{00}_{LL}=-g(t^{00}_{LL}+T^{00}).
\end{equation}
After some manipulations, it follows that
\begin{eqnarray}\label{43}
t^{00}_{LL}&=&-\frac{3}{4}(\frac{\alpha}{a})^2
\frac{(x^2+a^2y^2)(y^2+a^2x^2)}{(x^2+y^2)^2}
-\frac{1}{4}(\frac{\alpha}{a})^2\frac{(a^2-1)(x^2+a^2y^2)}{x^2+y^2}\nonumber\\
&-&\frac{x^3y^5(a^2-1)^3}{a^2e^{\alpha t}(x^2+y^2)^5}
-\frac{2x^4y^4(a^2-1)^3}{a^2e^{\alpha t}(x^2+y^2)^5}
+\frac{x^4y^2(a^2-1)^2(x^2+a^2y^2)}{a^2e^{\alpha t}(x^2+y^2)^5}\nonumber\\
&+&\frac{1}{2}(\frac{\alpha}{a})^2
\frac{x^2y^2(x^2+a^2y^2)(y^2+a^2x^2)(a^2-1)^2}{(x^2+y^2)^4}\nonumber\\
&-&\frac{1}{2}(\frac{\alpha}{a})^2
\frac{(x^2+a^2y^2)^2(y^2+a^2x^2)^2}{(x^2+y^2)^4}.
\end{eqnarray}
Also, $T^{00}$ is given by
\begin{equation}\label{44}
T^{00}=M\delta(x)\delta(y).
\end{equation}
Using these values in Eq.(\ref{42}), it follows that
\begin{eqnarray}\label{45}
\tau^{00}_{LL}&=&a^2e^{3\alpha t}[-\frac{3}{4}(\frac{\alpha}{a})^2
\frac{(x^2+a^2y^2)(y^2+a^2x^2)}{(x^2+y^2)^2}
-\frac{1}{4}(\frac{\alpha}{a})^2\frac{(a^2-1)(x^2+a^2y^2)}{x^2+y^2}\nonumber\\
&-&\frac{x^3y^5(a^2-1)^3}{a^2e^{\alpha t}(x^2+y^2)^5}
-\frac{2x^4y^4(a^2-1)^3}{a^2e^{\alpha t}(x^2+y^2)^5}
+\frac{x^4y^2(a^2-1)^2(x^2+a^2y^2)}{a^2e^{\alpha t}(x^2+y^2)^5}\nonumber\\
&+&\frac{1}{2}(\frac{\alpha}{a})^2
\frac{x^2y^2(x^2+a^2y^2)(y^2+a^2x^2)(a^2-1)^2}{(x^2+y^2)^4}\nonumber\\
&-&\frac{1}{2}(\frac{\alpha}{a})^2
\frac{(x^2+a^2y^2)^2(y^2+a^2x^2)^2}{(x^2+y^2)^4}+M\delta(x)\delta(y)].
\end{eqnarray}
Inserting all these values in Eq.(\ref{41}), we have
\begin{eqnarray}\label{46}
\tau^{00}&=&\frac{1}{2\kappa}[F(R_{0})a^2e^{3\alpha
t}\{-\frac{3}{4}(\frac{\alpha}{a})^2
\frac{(x^2+a^2y^2)(y^2+a^2x^2)}{(x^2+y^2)^2}\nonumber\\
&-&\frac{1}{4}(\frac{\alpha}{a})^2\frac{(a^2-1)(x^2+a^2y^2)}{x^2+y^2}
-\frac{x^3y^5(a^2-1)^3}{a^2e^{\alpha t}(x^2+y^2)^5}
-\frac{2x^4y^4(a^2-1)^3}{a^2e^{\alpha t}(x^2+y^2)^5}\nonumber\\
&+&\frac{x^4y^2(a^2-1)^2(x^2+a^2y^2)}{a^2e^{\alpha t}(x^2+y^2)^5}
-\frac{1}{2}(\frac{\alpha}{a})^2
\frac{(x^2+a^2y^2)^2(y^2+a^2x^2)^2}{(x^2+y^2)^4}\nonumber\\
&+&\frac{1}{2}(\frac{\alpha}{a})^2
\frac{x^2y^2(x^2+a^2y^2)(y^2+a^2x^2)(a^2-1)^2}{(x^2+y^2)^4}
+M\delta(x)\delta(y)\}\nonumber\\
&+&(F(R_{0})R_{0}-f(R_{0}))].
\end{eqnarray}

Now we discuss a well-known special case for the choice of $f(R)$
model \cite{024}
\begin{equation}\label{47}
f(R)=R-(-1)^{n-1}\frac{a}{R^{n}}+(-1)^{m-1}bR^{m},
\end{equation}
where $m$ and $n$ are positive integers. For $R\equiv
R_{0}=-4\Lambda$, we have
\begin{equation}\label{48}
f(R_0)=-4\Lambda+\frac{a}{(4\Lambda)^n}-b(4\Lambda)^m
\end{equation}
and
\begin{equation}\label{49}
f'(R_0)=\frac{(4\Lambda)^{n+1}+an+bm(4\Lambda)^{m+n}}{(4\Lambda)^{n+1}}.
\end{equation}
Inserting these values in Eq.(\ref{46}), it follows that
\begin{eqnarray}\label{50}
\tau^{00}&=&\frac{1}{2\kappa}[(\frac{(4\Lambda)^{n+1}
+an+bm(4\Lambda)^{m+n}}{(4\Lambda)^{n+1}})a^2e^{3\alpha
t}\{-\frac{3}{4}(\frac{\alpha}{a})^2\nonumber\\
&\times&\frac{(x^2+a^2y^2)(y^2+a^2x^2)}{(x^2+y^2)^2}
-\frac{1}{4}(\frac{\alpha}{a})^2\frac{(a^2-1)(x^2+a^2y^2)}{x^2+y^2}\nonumber\\
&-&\frac{x^3y^5(a^2-1)^3}{a^2e^{\alpha t}(x^2+y^2)^5} +
\frac{\alpha^2x^2y^2(x^2+a^2y^2)(y^2+a^2x^2)(a^2-1)^2}{2a^2(x^2+y^2)^4}\nonumber\\
&+&\frac{x^4y^2(a^2-1)^2(x^2+a^2y^2)}{a^2e^{\alpha t}(x^2+y^2)^5}
-\frac{\alpha^2(x^2+a^2y^2)^2(y^2+a^2x^2)^2}{2a^2(x^2+y^2)^4}\nonumber\\
&-&\frac{2x^4y^4(a^2-1)^3}{a^2e^{\alpha
t}(x^2+y^2)^5}+M\delta(x)\delta(y)\}\nonumber\\
&+&\frac{b(1-m)(4\Lambda)^{m+n}-a(1+n)}{(4\Lambda)^n}].
\end{eqnarray}
This model must satisfy the constant curvature condition given by
Eq.(\ref{06}). Imposing this condition, we obtain
\begin{equation}\label{51}
a(n+2)+b(m-2)(4\Lambda)^{m+n}=(4\Lambda)^{n+1}-2\kappa
M\delta(x)\delta(y)(4\Lambda)^n.
\end{equation}
For a particular case, when $m=2$ or $b=0$, it reduces to
\begin{equation}\label{52}
a=\frac{(4\Lambda)^{n+1}-2\kappa
M\delta(x)\delta(y)(4\Lambda)^n}{n+2}.
\end{equation}
It satisfies the constant scalar curvature condition which is
necessary for the acceptability of the model.

Now we discuss another important $f(R)$ model \cite{030} given by
\begin{equation}\label{53}
f(R)=R-a\ln(\frac{|R|}{k})+(-1)^{n-1}bR^{n}.
\end{equation}
In this case, the $00$-component of the generalized Landau-Lifshitz
EMC takes the form
\begin{eqnarray}\label{54}
\tau^{00}&=&\frac{1}{2\kappa}[(\frac{4k\Lambda-a+bkn(4\Lambda)^n}{4k\Lambda})a^2e^{3\alpha
t}\{-\frac{3}{4}(\frac{\alpha}{a})^2
\frac{(x^2+a^2y^2)(y^2+a^2x^2)}{(x^2+y^2)^2}\nonumber\\
&-&\frac{1}{4}(\frac{\alpha}{a})^2\frac{(a^2-1)(x^2+a^2y^2)}{x^2+y^2}
-\frac{x^3y^5(a^2-1)^3}{a^2e^{\alpha t}(x^2+y^2)^5}
-\frac{2x^4y^4(a^2-1)^3}{a^2e^{\alpha t}(x^2+y^2)^5}\nonumber\\
&+&\frac{x^4y^2(a^2-1)^2(x^2+a^2y^2)}{a^2e^{\alpha t}(x^2+y^2)^5}
-\frac{1}{2}(\frac{\alpha}{a})^2
\frac{(x^2+a^2y^2)^2(y^2+a^2x^2)^2}{(x^2+y^2)^4}\nonumber\\
&+&\frac{1}{2}(\frac{\alpha}{a})^2
\frac{x^2y^2(x^2+a^2y^2)(y^2+a^2x^2)(a^2-1)^2}{(x^2+y^2)^4}+M\delta(x)\delta(y)\}\nonumber\\
&+&a\ln(\frac{4\Lambda}{k})+\frac{a}{k}+b(1-n)(4\Lambda)^n].
\end{eqnarray}
Also, the constant scalar curvature condition gives
\begin{equation}\label{55}
2a\ln(\frac{4\Lambda}{k})-a+4\Lambda=b(4\Lambda)^n(n-2)+2\kappa
M\delta(x)\delta(y).
\end{equation}
For $n=2$ or $b=0$, this reduces to
\begin{equation}\label{56}
a=\frac{2\kappa
M\delta(x)\delta(y)-4\Lambda}{2\ln(\frac{4\Lambda}{k})-1}
\end{equation}
which satisfies the constant curvature condition necessary for the
acceptability of the model given by (\ref{53}).

\section{Summary and Conclusion}

In this paper, the well-posed problem of energy-momentum
localization has been discussed in the context of $f(R)$ gravity.
For this purpose, we use the generalized Landau-Lifshitz
energy-momentum complex. We evaluate energy density of some static
plane symmetric solutions by using this energy-momentum complex. The
energy density of the cosmic string spacetime is also calculated.
Further, this quantity is investigated for some important $f(R)$
models. We have mainly considered two types of models, one with
negative and positive powers of curvature and other including
logarithmic term of curvature. These models have been found
consistent with the solar system test and it has been shown that the
model with negative and positive power of curvature unifies
inflation and cosmic acceleration. The terms with positive powers of
curvature provide the inflationary stage while the terms with
negative powers of curvature serves as an alternative for dark
energy which is responsible for cosmic acceleration.

The model with logarithm term is also cosmologically important as it
suggests due the logarithmic term which may be responsible for the
current acceleration of the universe. It has been shown that the
chosen $f(R)$ models satisfy the constant scalar curvature condition
which is the necessary requirement for the validity of these models.
We have also explored the stability condition for these models. The
results (\ref{22}), (\ref{27}) and (\ref{32}) show that the energy
density expressions are well-defined in these cases. These results
can reduce to GR by taking $f(R)=R$ in all the cases. We would like
to mention here that we have calculated for the first time, to our
knowledge, the energy density for a non-vacuum case (cosmic string
spacetime) using the generalized Landau-Lifshitz energy-momentum
complex.

This work adds some knowledge about the longstanding and crucial
problem of the localization of energy. It gives the energy density
expressions for different solutions with important $f(R)$ models
which may help at some stage to overcome the theoretical
difficulties in the cosmological and astrophysical context. It would
be interesting to find the Landau-Lifshitz EMC for non-constant
scalar curvature. The extension of other EMCs in the context of
$f(R)$ gravity as well in other versions of $f(R)$ gravity would
also be worthwhile.

\end{document}